# Large Spin Nernst Effect in Ni$_{70}$Cu$_{30}$ Alloy

Wen-Yuan Li[1,2], Chia-Hsi Lin[1], Guang-Yu Guo[1,3], Ssu-Yen Huang[1,4] and Danru Qu[2,4*]

[1]*Department of Physics, National Taiwan University, Taipei 10617, Taiwan*

[2]*Center for Condensed Matter Sciences, National Taiwan University, Taipei 10617, Taiwan*

[3]*Physics Division, National Center for Theoretical Sciences, Taipei 10617, Taiwan*

[4]*Center of Atomic Initiatives for New Materials, National Taiwan University, Taipei 10617, Taiwan*

[*]danru@ntu.edu.tw (D.Q.)

Abstract:

The interplay among heat, spin, and charge is the central focus in spin caloritronic research. While the longitudinal heat-to-spin conversion via the spin Seebeck effect has been intensively studied, the transverse heat-to-spin conversion via the spin Nernst effect (SNE) has not been equally explored. One major challenge is the minuscule signals generated by the SNE, which are often mixed with the background noises. In this work, we overcome this difficulty by studying the thin films of Ni$_{70}$Cu$_{30}$ alloy with not only a sizable spin Hall angle but also a large Seebeck coefficient. We observe in the Ni$_{70}$Cu$_{30}$ alloy a large spin Nernst effect with an estimated spin Nernst angle ($\theta_{SN}$) ranging from -28% to -72%. In comparison, the $\theta_{SN}$ for Pt is -8.2%. Our *ab initio* calculation reveals that the large spin Nernst conductivity ($\alpha_{SN}$) in Ni$_{70}$Cu$_{30}$ is caused by the Fermi energy shift to the steepest slope of the spin Hall conductivity curve due to electron doping from 30% Cu. Our study provides critical directions in searching for materials with a large spin Nernst effect.

Spin-caloritronics explores the interactions among spin, charge, and heat. It has significant potential for energy harvesting by converting heat into spin information and further electrical energy. While most studies focus on the longitudinal heat-to-spin conversion via the spin Seebeck effect (SSE) in ferromagnetic insulators (FMI) [1], the transverse heat-to-charge conversion via the anomalous Nernst effect (ANE) in ferromagnetic metals (FM) [2], and the transverse charge to spin interconversion via the spin Hall effect (SHE) in non-magnetic metals (NM) [3-4]; their counterpart, the spin Nernst effect (SNE), that converts heat to transverse spin current in non-magnetic metals [5-10], has not been adequately discussed. To date, only two 5d metals, platinum (Pt) and tungsten (W), are experimentally reported to exhibit SNE [5-9]. Compared to the wide variety of materials reported to achieve sizable SHE, the SNE materials capable of generating transverse spin current in response to heat current remain seriously underexplored.

In the SHE, as shown in Fig. 1(a), a charge current with density $j_C$ flows through a non-magnetic metal with spin-orbit coupling, generates a transverse spin current with density $j_S$, described as [4]

$$j_S = \theta_{SH} \frac{\hbar}{2e} j_C \times \sigma \qquad (1)$$

In this equation, $\theta_{SH}$ represents the spin Hall angle, which characterizes the efficiency of charge and spin interconversions, $\sigma$ is the spin index.

Similarly, in the SNE, a transverse spin current with density $j_S$ is generated under an inplane temperature gradient $\nabla T$ applied in an NM, as shown in Fig. 1(b), and is described as [5-10]

$$j_S = -\theta_{SN} \frac{\hbar}{2e} \frac{S}{\rho} \nabla T \times \sigma \qquad (2)$$

In this equation, $\theta_{SN}$ represents the spin Nernst angle, which characterizes the efficiency of converting heat flow to spin current. $S$ is the Seebeck coefficient of the material, and $\rho$ is the resistivity.

To detect the spin current generated from SNE, one could utilize an experimental geometry with an NM/FMI bilayer structure, similar to the spin Hall magnetoresistance (SMR) measurement [11], as shown in Fig. 1(c) and (e). In the SMR measurement, $J_C$ induces a transverse $J_S$ in the NM, which then flows vertically into the FMI layer with magnetization $M$. Depending on the relative orientations between the spin index $\sigma$ and $M$, $J_S$ is partially reflected at the interface or absorbed by the FM metal, as shown in Fig. 1(c) and (e), respectively. The reflected spin current further converts back to a charge current via ISHE [12] and results in a change of longitudinal resistivity $\Delta\rho = \rho_y - \rho_x$. Here, $\rho_x$ and $\rho_y$ are denoted as the resistivity $\rho$ obtained under $H_x$ and $H_y$, respectively. By the spin drift-diffusion model [13], the SMR ratio is characterized as

$$\frac{\Delta\rho}{\rho} = -\theta_{SH}^2 \frac{\lambda_S}{t} \frac{2\rho\lambda_S G_{\uparrow\downarrow} \tanh^2 \frac{t}{\lambda_S}}{1 + 2\rho\lambda_S G_{\uparrow\downarrow} \coth^2 \frac{t}{\lambda_S}} \qquad (3)$$

Here, $\lambda_S$, and $t$ are the spin diffusion length and thickness of the NM metal, respectively. $G_{\uparrow\downarrow}$ denotes the spin mixing conductance of the NM/FMI interface. $\theta_{SH}$ denotes the spin Hall angle that measures the conversion between charge and spin.

Similarly, when we replace $J_C$ with $\nabla T$, as shown in Fig. 1(d) and (f), changes of the longitudinal voltage $\Delta V$, thus the change of the Seebeck coefficient, $\Delta S = S_y - S_x$, are expected due to the heat, spin, and charge interconversions induced by SNE and ISHE. Here, $S$ is the Seebeck coefficient defined as $S = -\Delta V/\Delta T$. $S_x$ and $S_y$ are denoted as the Seebeck coefficient $S$ obtained under $H_x$ and $H_y$, respectively. This voltage is caused by the spin Nernst induced magneto-thermopower (SNMT) [5]. The SNMT signal is thus characterized by a modified version of eq. (3),

$$\frac{\Delta S}{S} = \theta_{SH}\theta_{SN} \frac{\lambda_S}{t} \frac{2\rho\lambda_S G_{\uparrow\downarrow} \tanh^2 \frac{t}{\lambda_S}}{1 + 2\rho\lambda_S G_{\uparrow\downarrow} \coth^2 \frac{t}{\lambda_S}} \qquad (4)$$

In this equation, $S$ is the Seebeck coefficient of the material, $\Delta T$ is the temperature difference, $\theta_{SN}$ is the spin Nernst coefficient, including the heat current generated and Seebeck voltage generated transverse spin current [14].

Although the SNE is a viable route for generating the spin current, the detection of SNE is challenging. Essentially, $\nabla T$ drives the motion of charge carriers in the NM. A typical

Seebeck coefficient for a material, for example, $S_{Cu}$ is 1.8 μV/K at room temperature (RT) [15]. For a temperature difference of 10 K, the Seebeck voltage $\Delta V = V_H - V_C$, generated in Cu, is only -18 μV. Here, $V_H$ and $V_C$ are voltages obtained from the hot and cold sides, respectively. For SHE, to enhance the generated spin current, one could increase the current density as large as 6 orders of magnitude, for example, from 1 μA to 1 A. However, it is impractical to significantly increase a temperature difference, for example, from 10 K to 1000 K, by 2 orders of magnitude. Thus, when environmental noise obscures the SNMT voltage, simply increasing the temperature difference may not be a practical solution.

According to Eq. (4), the SNMT signal strongly depends on the spin Hall angle $\theta_{SH}$ and Seebeck coefficient $S$. Thus, a material with significant $\theta_{SH}$ and, most importantly, a large $S$ would be ideal to capture the SNE. The NiCu alloys are such materials with sizable $\theta_{SH}$ [16] and large $S$. It is reported that the band structure of non-magnetic Ni is similar to the band structure of Pt [17]. This ensures the sizable spin-to-charge interconversion. Moreover, NiCu alloys are widely used in thermocouples due to their large $S$, such as the T, J, and E-type thermocouples, where a Seebeck voltage is developed upon a heat gradient. Combining these advantages, we choose the $Ni_{70}Cu_{30}$ alloy to study the SNE, which remains non-magnetic at room temperature for films thinner than 10 nm. From the SMR and SNMT measurements, we surprisingly observe a large spin Nernst angle of -71.7% for the $Ni_{70}Cu_{30}$ alloy, which is nearly 9 times larger than that for Pt. The *ab initio* calculations further reveal that with 30% Cu doping, the Fermi energy for the nonmagnetic NiCu alloy sits at the steepest slope of the spin Hall conductivity plot, which gives the maximal spin Nernst conductivity for $Ni_{70}Cu_{30}$. Our study not only adds a superior collection to the spin Nernst materials but also provides critical directions in searching for potential materials with significant SNE.

In our study, we fabricate the $Ni_{70}Cu_{30}$ thin films using magnetron sputtering on an alloy target with the same stoichiometry. The $Ni_{70}Cu_{30}$ alloy films are deposited onto 0.5 mm thick polycrystalline yttrium iron garnet ($Y_3Fe_5O_{12}$, YIG) substrates at room temperature without heat treatment. For the SMR and SNMT measurements, the $Ni_{70}Cu_{30}$ films are subsequently patterned into Hall bar structures using photo-lithography and $Ar^+$ ion etching, as shown in Fig. 2(a) and (b). The Hall bar width is 100 μm, and the voltage length is 3 mm. For spin Seebeck measurement, we use continuous films without lithography, as shown in Fig. 3(a). All measurements are performed at room temperature.

We first measure the SMR for the 7-nm thick $Ni_{70}Cu_{30}$ film on YIG. We perform the SMR measurement using the four-probe technique. As shown in Fig. 2(a), a charge current of 1 μA is applied along the $x$-axis, and a longitudinal voltage is measured under an external magnetic field applied along the $x$- or $y$-axis, denoted as $H_x$ and $H_y$, respectively. The resistivity $\rho$ obtained under $H_x$ and $H_y$ are denoted as $\rho_x$ and $\rho_y$, respectively. We average the signals using the symmetric relation $\rho_{sym}(H)=[\rho(H)+\rho(-H)]/2$. Since the 7-nm thick $Ni_{70}Cu_{30}$ is non-magnetic, as demonstrated in supplementary material S1 [18], the anisotropic-like magnetoresistance (MR), that depends on the YIG magnetization, as shown in Fig. 2(c), with $\rho_x > \rho_y$, can be solely attributed to the pure spin current generation, reflection, and conversion in Ni$_{70}$Cu$_{30}$. From the results in Fig. 2(c), we obtain $\rho = 61.8 \pm 1.0$ μΩ·cm and $\Delta\rho = (7.4 \pm 2.4) \times 10^{-4}$ μΩ·cm at the saturation magnetic field $H$ = - 500 Oe, for the 7-nm thick Ni$_{70}$Cu$_{30}$ on YIG. The uncertainty accounts for the linear drift of resistivity on large magnetic field. Thus, the spin Hall magnetoresistance (SMR) ratio is $(1.2 \pm 0.4) \times 10^{-5}$. According to Eq. (3), by considering $\lambda_S = 1.8$ nm, $G_{\uparrow\downarrow} = 7.3 \times 10^{13}$ Ω$^{-1}$m$^{-2}$ [16], we obtain a $\theta_{SH}$ for Ni$_{70}$Cu$_{30}$ as $1.9 \pm 0.6$ %. The value of $\lambda_S$ is based on the spin Seebeck results discussed later, while the assumed value of $G_{\uparrow\downarrow}$ is based on the comparable crystallinity (polycrystalline, face-centered-cubic structure), for room temperature sputtered Cu, Ni and NiCu alloy on Si [16] and YIG (see Supplementary Materials [18]).

As a reference, we fabricate 3-nm thick Pt thin films on YIG since the SHE in Pt has been extensively studied and its SNE has been investigated [5]. The resistivity for the 3-nm thick Pt is $73.9 \pm 1.3$ μΩ·cm. An SMR with $\rho_x > \rho_y$ is observed for Pt on YIG, as shown in Fig. 2(e). The SMR ratio of $(2.10 \pm 0.04) \times 10^{-4}$ is about an order larger compared to that of Ni$_{70}$Cu$_{30}$. By using $\lambda_S = 1.5$ nm and $G_{\uparrow\downarrow} = 4 \times 10^{14}$ Ω$^{-1}$m$^{-2}$ for Pt/YIG [5], we obtain its $\theta_{SH}$ to be about $4.0 \pm 0.1$ %.

Next, we study the SNE via the magneto-thermopower. The SNMT is measured by probing the longitudinal voltage with a heat flow driven along the $x$-axis, as shown in Fig. 2(b).

The external field is also applied along the *x*- and *y*-axes, with voltages obtained, denoted as $V_x$ and $V_y$, respectively. An in-plane temperature gradient $\nabla T$ of 1.5 K/mm is applied by placing the short edges of the 12-mm long and 4-mm wide sample on top of two aluminum blocks: one side as a heat source attached with resistive heaters, which is thermally isolated from the sample holder, and the other as a heat sink, which is thermally glued to a Cu plate, as shown in Fig. S2(a). The temperature on both sides is measured using *K*-type thermocouples.

On the other hand, we also note that a parasitic vertical $\nabla T$ cannot be eliminated entirely when an inplane $\nabla T$ is applied [5, 7], as shown in Fig. S3(b). Under a vertical $\nabla T$, a longitudinal magnonic spin current is generated in the FMI through the spin Seebeck effect (SSE) [1], as shown in the inset of Fig. 4(b). With the NM/FMI bilayer structure, an SSE/ISHE voltage could also present in addition to the SNMT, as shown in Fig. S3(c). However, the SSE and SNMT voltages have odd and even symmetry on *H*, respectively. Thus, to obtain the SNMT signal, we separate the SSE and SNMT signals by the following symmetric operations:

$$V_{sym}(H) = (V(+H) + V(-H))/2 \qquad (5)$$

$$V_{asym}(H) = (V(+H) - V(-H))/2 \qquad (6)$$

where $V_{sym}(H)$ and $V_{asym}(H)$ denote symmetric SNMT and anti-symmetric SSE contributions, respectively. By reversing the magnetic field, we are able to construct a complete hysteresis loop containing the forward and backward field scans.

As shown in Fig. 2(d), a symmetric signal containing the SNMT contributions is separated from the asymmetric SSE contributions based on Eq. (5). The NiCu alloy has a positive Seebeck voltage $V_S$ about 90.8 μV under $\Delta T = 7.3 \pm 0.1$ K. Notice that this value contains a contribution from the Seebeck voltage of the wires in our measurement setup, as discussed in Supplementary S4 [18]. Taking the Seebeck effect from the bonding wires into account, as $V_S = -(S_{\text{NiCu}} - S_{\text{wire}})\Delta T$, with $S_{\text{wire}} = 1.7 \pm 0.1$ μV/K, we obtain an $S_{\text{NiCu}}$ for the 7-nm thick NiCu as $-10.7 \pm 0.2$ μV/K. More interestingly, an SMR-like SNMT signal is captured with $V_x > V_y$ and $\Delta V = V_y - V_x = -0.036 \pm 0.003$ μV at the saturation field $H =$ -500 Oe. We calculate the thermopower ratio $\Delta S/S_{\text{NiCu}} = -(4.6 \pm 0.3) \times 10^{-4}$, where $\Delta S = -(V_y-V_x)/\Delta T$. We find the SNMT ratio is about 38 times larger than its SMR ratio, indicating a higher heat-to-spin conversion efficiency with $\theta_{\text{SN}} = -71.7 \pm 23.2$ %, according to Eq (3) and (4). For thinner 5-nm-thick $Ni_{70}Cu_{30}$ film, as demonstrated in Supplementary S5 [18], its $\theta_{\text{SH}} \sim 2.2\%$, and $\theta_{\text{SN}} \sim -28\%$. From the negligible anomalous Hall signal for the 7-nm thick $Ni_{70}Cu_{30}$/YIG (See Supplementary S1), and a smaller $\theta_{\text{SN}}$ for thinner film, we consider the contribution from the interfacial magnetic proximity effect to the SMR and SNMT is negligible.

As a comparison, in the case of 3-nm thick Pt, a negative Seebeck voltage of about −10.5 μV is captured under the same $\Delta T = 7.3 \pm 0.1$ K. The difference of the Seebeck voltage for $M$ saturates along $x$ and $y$ is $0.010 \pm 0.001$ μV. Considering the thermovoltage contribution from the Cu bonding wire to the total signal, we obtain an $S_{Pt}$ for the 3-nm thick

Pt as $3.1 \pm 0.1$ μV/K. Thus, the SNMT ratio for Pt is $\Delta S/S_{\text{Pt}} = (-4.4 \pm 0.5) \times 10^{-4}$, leading to a $\theta_{SN} = -8.2 \pm 0.9$ %, which is about 9 times smaller than that of Ni$_{70}$Cu$_{30}$.

We point out here that the Seebeck voltages from the bonding wires connecting to the voltmeter need to be carefully treated, especially for materials like Pt and W with small S. This essential point, however, has been ignored in previous research. When one side of the bonding wires is connected to the sample edges at high or low temperatures, and the other side is connected to a voltmeter at RT, a Seebeck voltage is developed in the wires, contributing to the total thermal voltage. When the SNE material has small $S$, such as Pt, without considering the bonding Cu wires, from Fig. 2(f), we may falsely obtain an $S$ for Pt as 10.5 μV / 7.3 K = 1.4 μV/K, which leads to a wrong estimation of $\theta_{SN} = -17.3$ %. While the revised $\theta_{SN}$, accounting for the Seebeck effect in the wires, has a value of - 8.2 %. Thus, when studying the spin Nernst effect via SNMT or any thermally induced voltages, the $S$ of the connecting wires needs to be treated carefully, the overlook of which will result in an incorrect estimation of the $\theta_{SN}$.

We understand the large response in the transverse heat-to-spin conversion for Ni$_{70}$Cu$_{30}$ in analogy to the anomalous Nernst effect (ANE) [19-22]. In ANE, the transverse Seebeck coefficient $S_{xy}$ is proportional to the transverse anomalous Nernst conductivity (ANC) $\alpha_{xy}$ via

$$S_{xy} = \rho(\alpha_{xy} - S_{xx}\sigma_{xy}) \quad (7)$$

$$\alpha_{xy} = \frac{\pi^2 k_B^2 T}{3e} \left(\frac{\partial \sigma_{xy}}{\partial E}\right)_{E_F} \qquad (8)$$

Here, $\sigma_{xy}$ is the anomalous Hall conductivity (AHC), $k_B$ is the Boltzmann constant, $e$ is the electric charge, and $E_F$ is the Fermi energy. Eq. (8) is the Mott relation. From the equation, we know that the $\alpha_{xy}$ depends on the derivative of the $\sigma_{xy}$ plot. Similarly, the spin Nernst effect essentially probes the spin Hall conductivity (SHC) change at Fermi surface. A large spin Nernst conductivity (SNC) requires a deep slope in the plot of the SHC ($\sigma_{SH}$). Compared to a single element Pt or W, by altering the ratio of the NiCu alloy, we could change the Fermi level and achieve a significant $\alpha_{SN}$.

This scenario is firmly supported by our *ab initio* calculations of the SHC $\sigma_{SH}$ and SNC $\alpha_{SN}$ as a function of energy for the non-magnetic Ni with face-centered-cubic (fcc) crystal structure (see supplementary material S6 [18] for the computational details). Fig. 3(a) shows that pure non-magnetic Ni has a large SHC of about 1600 $(\hbar/e)$(S/cm), while the $Ni_{70}Cu_{30}$ alloy has a slightly decreased value in the range of 1150 $(\hbar/e)$(S/cm) due to the small upwards shift of the Fermi level (by merely 0.088 eV). Excitingly, since the Fermi energy of $Ni_{70}Cu_{30}$ now sits at the sharpest slope in the $\sigma_{SH}$-energy plot, the $Ni_{70}Cu_{30}$ alloy has a maximal $\alpha_{SN}$ of 3.5 $(\hbar/e)$(A/m-K), as shown in Fig. 3(b), as expected from the Mott relation in Eq. (8). Interestingly, Fig. 3(c) shows that the Fermi level of the $Ni_{70}Cu_{30}$ alloy is allocated at the energy position where the density of states (DOS) also has a steep slope [see Fig. 3(c)], and this explains the large observed negative Seebeck coefficient in $Ni_{70}Cu_{30}$. In our experiment, we have $\rho_{xx\text{-NiCu}}$ = 61.8 μΩcm, $\sigma_{xx\text{-NiCu}}$ = 16200 (S/cm), $\theta_{SH\text{-NiCu}}$ = 1.9 ± 0.6 %, thus an estimated SHC $\sigma_{SH\text{-NiCu}} \approx (\hbar/2e)\theta_{SH}\sigma_{xx}$ = 150 ± 47 ($\hbar/e$)(S/cm). Similarly, we have $S_{xx\text{-NiCu}}$ = - 10.7 μV/K, $\theta_{SN\text{-NiCu}}$ = - 71.7 ± 23.2 %, thus an estimated SNC $\alpha_{SN\text{-NiCu}} = (\hbar/2e)\theta_{SN}S_{xx}\sigma_{xx}$ = 6.2 ± 2 ($\hbar/e$)(A/m-K). Overall, our experimental results qualitatively follow the theoretical predictions, while the differences could be due to the finite size effect from thin films, the further up-shift of the Fermi energy, or the extrinsic contributions caused by Cu doping, which are not included in the rigid band model. These values are summarized in Table I.

We further study the spin Seebeck effect to verify the spin Hall angle for Ni$_{70}$Cu$_{30}$. In the SSE measurement, a temperature gradient is applied along the $z$-axis by sandwiching the sample between a heater and a heat sink. The temperature above and below the sample is measured using $K$-type thermocouples. The external magnetic field is applied along the $y$-axis, and the ISHE voltage is probed along the $x$-axis. In contrast to the symmetric SMR and SNMT signals, the SSE signals are asymmetric in a magnetic field, as shown in Fig. 4(a). The ISHE voltage $V_{ISHE}$ is described as [23]:

$$\Delta V_{ISHE} = 2C\rho l \nabla T \theta_{SH} \frac{\lambda_S}{t} \tanh \frac{t}{2\lambda_S} \qquad (9)$$

Here, $l$ is sample length, $C = \frac{\gamma \hbar \rho' k_m^3 l_m}{4\pi M \pi^2} \frac{B_1 B_S}{B_2} k_B g_{eff}^{\uparrow\downarrow}$ is the spin injection coefficient, which depends on the spin mixing conductance $g_{eff}^{\uparrow\downarrow} = \frac{h}{e^2} G_{\uparrow\downarrow} = 1.9 \times 10^{18}$ m$^{-2}$, the magnetization $4\pi M$ = 1750 G, gyromagnetic ratio $\gamma = 1.48 \times 10^{11}$ s$^{-1}$T$^{-1}$, magnon diffusion length $l_m$ = 70 nm, finite ferromagnetic insulator thickness factor $\rho´$ ~1, maximum wavenumber $k_m = 2 \times 10^9$ m$^{-1}$, and parameters from the diffusion equation $B_1 = 0.55$, $B_s = 2.2 \times 10^{-4}$, $B_2 = 5.1 \times 10^{-3}$ [16], the electron charge $e$, the Boltzmann constant $k_B$ and the reduced Planck constant $\hbar$. Based on these parameters, $C = 11.9$ Am$^{-1}$K$^{-1}$. We measure several samples with different thicknesses, as shown in Fig. 4(b). By taking $\Delta T = 13.6 \pm 0.1$ K, and the sample size $l = 6 \pm 1$ mm, we fit the data with Eq. (9) and we obtain $\theta_{SH} = 1.4 \pm 0.3\%$ and $\lambda_S = 1.8 \pm 0.9$ nm for Ni$_{70}$Cu$_{30}$. This result is consistent with the $\theta_{SH}$ obtained from the SMR.

From these results, we summarize the essential criteria to search for materials with large spin Nernst effect: Firstly, a large Seebeck effect to enhance the heat-to-charge conversion; Secondly, an alloy material to tune Fermi energy to the highest SNC value; Thirdly, for the detection of SNE using SNMT method, a large spin Hall angle for the effective conversion of generated spin current back to the charge current. Excitingly, NiCu alloy meets all these criteria. Through SMR, SNMT, SSE measurements, and *ab initio* calculations, we obtain a large spin Nernst angle for Ni$_{70}$Cu$_{30}$ ranging from -28% to -72%. The significant SNC in NiCu is understood by the Fermi energy shift to the steepest slope of the SHC curve due to electron doping from the 30% Cu. We also point out that the seemingly trivial Seebeck effect of the connecting wires must be carefully treated to obtain a correct $\theta_{SN}$. Our study provides an essential cornerstone in studying and utilizing the spin Nernst effect.

Acknowledgement

We acknowledge support by the National Science and Technology Council under Grant No. NSTC 110-2112-M-002-047-MY3, No. NSTC 113-2628-M-002-019, No. NSTC 110-2123-M-002-010, No. NSTC 112-2123-M-002-001, and No. NSTC 113-2123-M-002-015, Center of Atomic Initiative for New Materials (AI-MAT), National Taiwan University, within Higher Education Sprout Project by the Ministry of Education in Taiwan, and Center for Quantum Science and Engineering (CQSE), National Taiwan University.

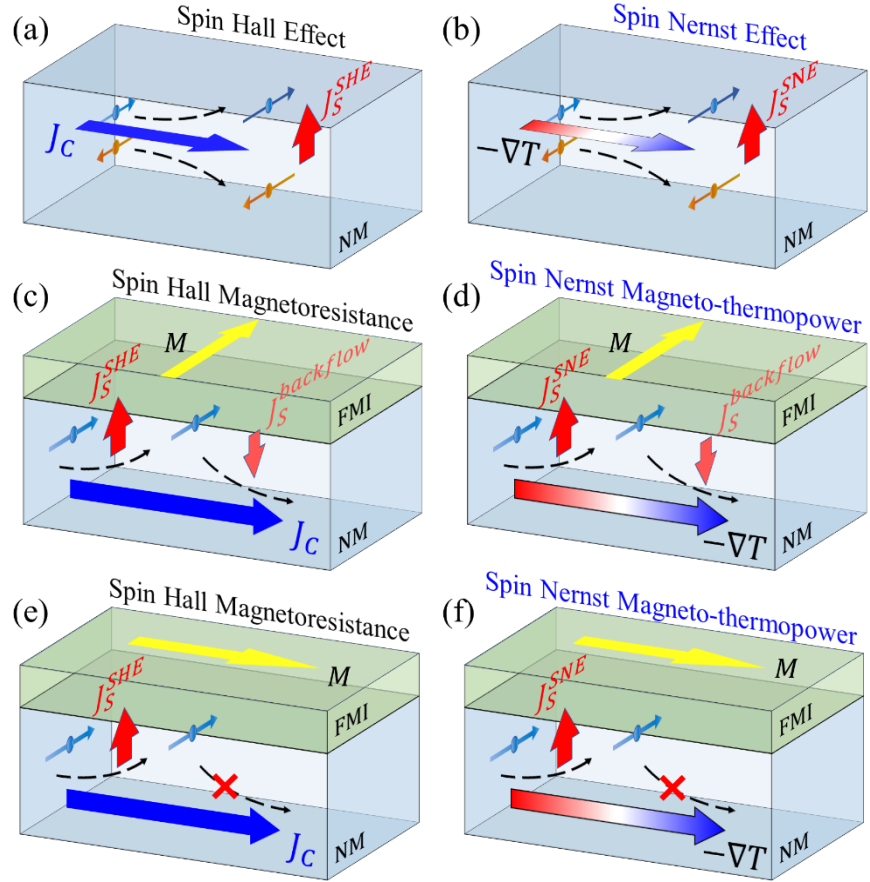

Fig. 1. Schematic illustrations of spin current phenomena and detection methods. (a) Spin Hall effect (SHE) in nonmagnetic metals (NM), where the spin current (red arrow), spin index (blue and orange arrows), and electric current (blue arrow) are mutually perpendicular. (b) Spin Nernst effect (SNE) in NM, with a similar configuration to SHE but with a temperature gradient (depicted with a French flag arrow) replacing electric current. Spin Hall magnetoresistance (SMR) in NM/FMI bilayer structures with magnetization (yellow arrow) (c) parallel and (e) perpendicular to spin orientation. Spin Nernst Magneto-thermopower (SNMT) in NM/FMI bilayer structures with magnetization (d) parallel and (f) perpendicular to spin orientation.

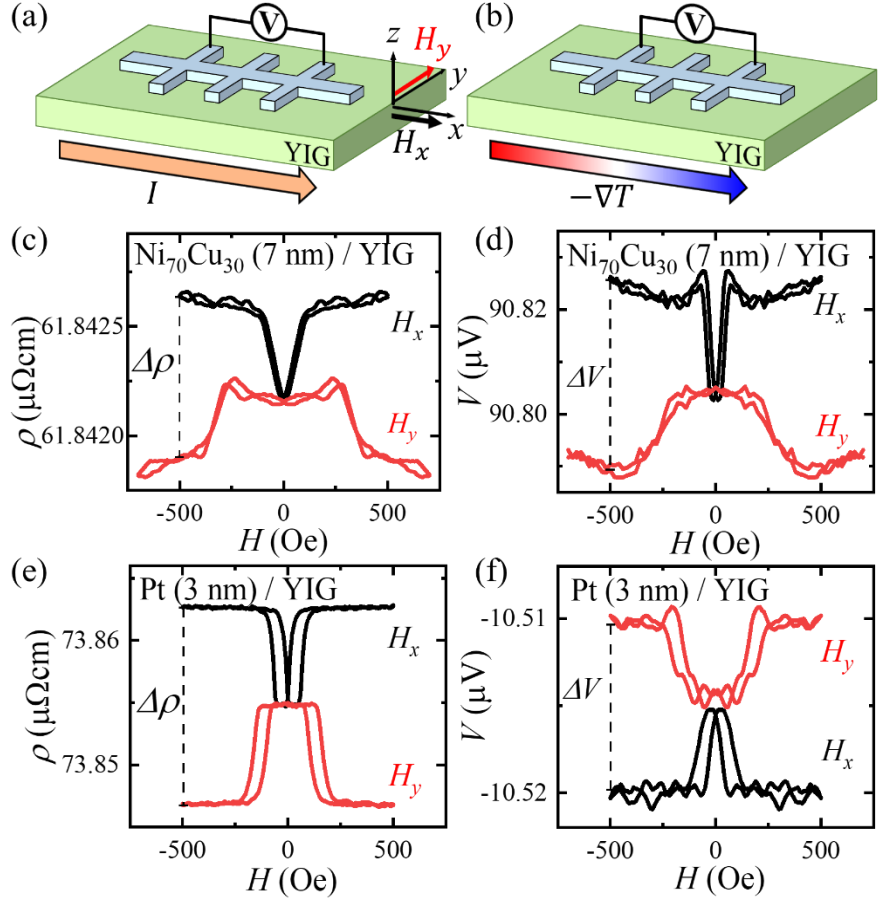

Fig. 2 Measurements of the spin Hall magnetoresistance (SMR) and spin Nernst magneto-thermopower (SNMT). In the schematic illustrations of (a) SMR and (b) SNMT measurement, the charge current $I$ and the temperature gradient $\nabla T$, respectively, are applied along $x$-axis, with voltage $V$ measured in the same direction. The magnetic field $H$ is applied and scanned along both $x$- and $y$-axis. (c) SMR and (d) SNMT results for $Ni_{70}Cu_{30}$ (7 nm) / YIG sample. (e) SMR and (f) SNMT results for Pt (3 nm)/ YIG sample. In each plot, black and red curves indicate measurements taken with the magnetic field aligned along the $x$- and $y$-axis, respectively. The data has been averaged by symmetric operations. A full hysteresis loop is constructed based on the average.

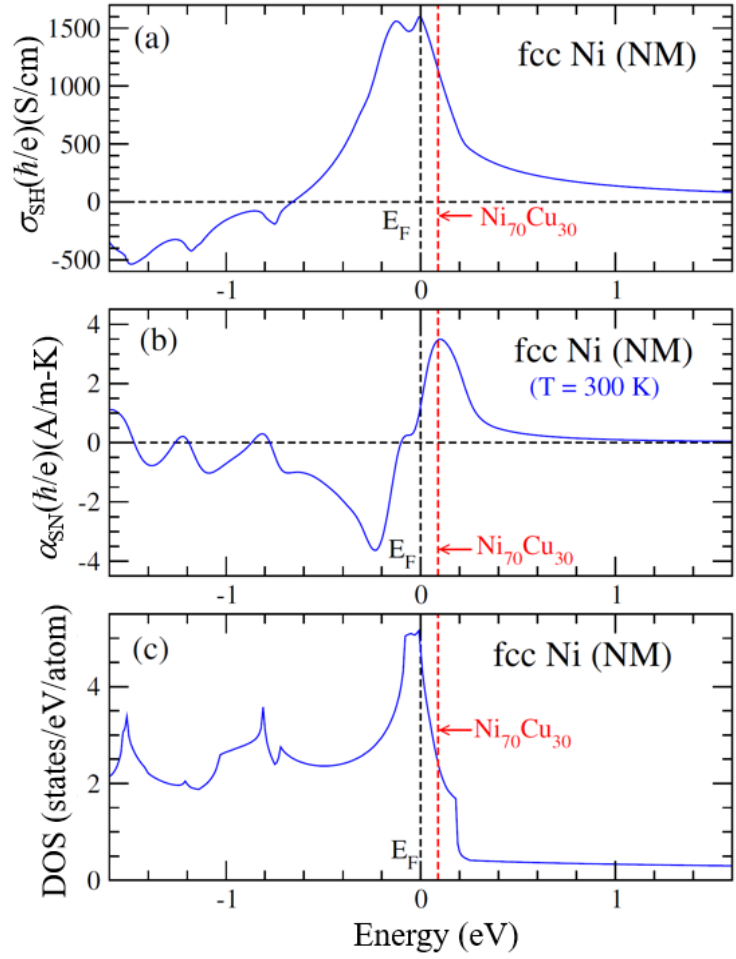

Fig. 3 *Ab initio* calculations based on face-centered-cubic (fcc) nonmagnetic (NM) Ni. (a) Spin Hall conductivity ($\sigma_{SH}$) and (b) spin Nernst conductivity ($\alpha_{SN}$) as well as (c) density of states (DOS) of fcc non-magnetic Ni from the *ab initio* calculation. The Fermi energy (at 0 eV) for pure Ni is marked by the black vertical dashed line. The red vertical dashed line denotes the Fermi level of the $Ni_{70}Cu_{30}$ alloy determined within the rigid band approximation (See supplementary note 4).

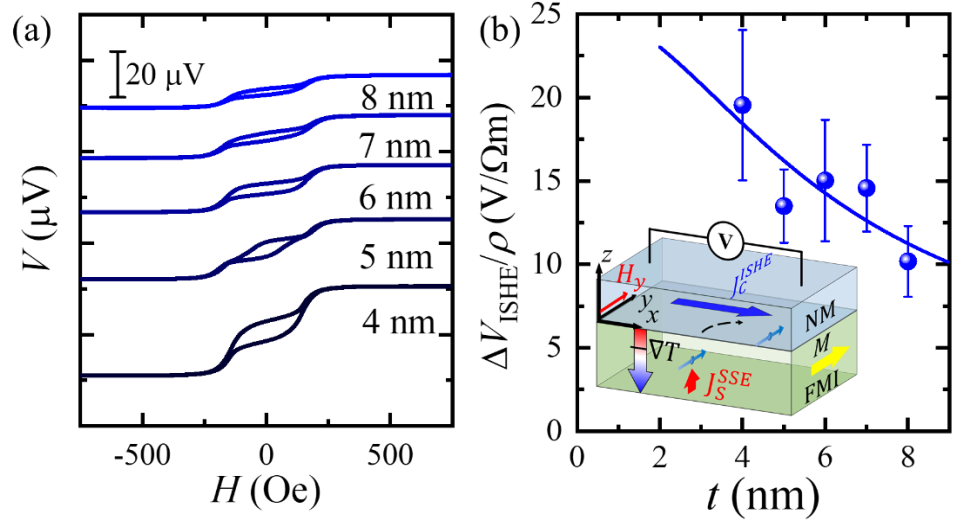

Fig. 4 Spin Seebeck measurements. (a) Field-dependent SSE/ISHE results for $Ni_{70}Cu_{30}$ layers varying from 4 to 8 nm. (b) Thickness-dependent SSE/ISSE results. Here, $t$ represents $Ni_{70}Cu_{30}$ thicknesses. Inset is schematic illustration of SSE measurements. The temperature gradient $\nabla T$ is applied along $z$-axis, with voltage $V$ measured in $x$-direction. The magnetic field $H$ is applied and scanned along $y$-axis.

| System | $\rho$ ($\mu\Omega$cm) | $\Delta\rho/\rho$ ($10^{-5}$) | $S$ ($\mu$V) | $\Delta S/S$ ($10^{-4}$) | $\theta_{SH}$ (%) | $\theta_{SN}$ (%) | $\sigma_{SH}^{EXP}$ ($\hbar/e$)(S/cm) | $\sigma_{SH}^{TH}$ | $\alpha_{SN}^{EXP}$ ($\hbar/e$)(A/m-K) | $\alpha_{SN}^{TH}$ |
|---|---|---|---|---|---|---|---|---|---|---|
| Ni$_{70}$Cu$_{30}$ (7 nm) | 61.8 ± 1.0 | 1.20 ± 0.4 | -10.7 ± 0.2 | -4.6 ± 0.3 | 1.9 ± 0.6 | -71.7 ± 23.2 | 150 ± 47 | 1153 Cal. | 6.2 ± 2 | 3.5 Cal. |
| Pt (3 nm) | 73.9 ± 1.3 | 21.0 ± 0.4 | 3.1 ± 0.1 | 4.4 ± 0.5 | 4.0 ± 0.1 | -8.2 ± 0.9 | 270 ± 10 | 2139 [22] | -0.17 ± 0.01 | -1.11 [22] |

TABLE I. Summarization of resistivity $\rho$, resistivity ratio $\Delta\rho/\rho$, thermopower $S$, thermopower ratio $\Delta S/S$, spin Hall angle $\theta_{SH}$, spin Nernst angle $\theta_{SN}$, spin Hall conductivity obtained experimentally $\sigma_{SH}^{EXP}$ and theoretically $\sigma_{SH}^{TH}$, and spin Nernst conductivity obtained experimentally $\alpha_{SN}^{EXP}$ and theoretically $\alpha_{SH}^{TH}$, for 7-nm thick Ni$_{70}$Cu$_{30}$ and 3-nm thick Pt films.